\documentclass{iau}
\usepackage{graphicx}
\usepackage{color,soul}

\title[Dead Zones in Disk Simulations] 
{Global Protoplanetary Disk Simulations:\\
   Dead Zone Formation and FUor Outbursts}

\author[Kundan Kadam, E. Vorobyov, et al.]   
{Kundan Kadam$^{1,*}$, E. Vorobyov$^{2,3}$, Zs. Reg\'{a}ly$^{1}$, \'{A}. K\'{o}sp\'{a}l$^{1,4}$, \\ 
P.  \'{A}br{\'a}ham$^{1}$}

\affiliation{$^1$Konkoly Observatory, MTA CSFK, Budapest, Hungary; \\ {$^*$}email: {\tt kundan.kadam@csfk.mta.hu} \\[\affilskip]
{$^2${The University of Vienna, Vienna, Austria}; $^3${Southern Federal University, Rostov-on-Don, Russia}; $^4${Max Planck Institute for Astronomy, Heidelberg, Germany}}.}

\pubyear{2019}
\volume{345}  
\jname{Origins: from the Protosun to the First Steps of Life}
\editors{Bruce G. Elmegreen, L. Viktor T\'oth, Manuel G\"udel, eds.}
\begin{document}

\maketitle

\begin{abstract}
We conducted global hydrodynamic simulations of protoplanetary disk evolution with an adaptive Shakura-Sunyaev $\alpha$ prescription to represent the layered disk structure, and starting with the collapse phase of the molecular cloud.
With the canonical values of model parameters, self-consistent dead zones formed at the scale of a few au. The instabilities associated with the dead zone and corresponding outbursts, similar to FUor eruptions, were also observed in the simulations.

\end{abstract}

\firstsection 
\section{Introduction}

T Tauri-type stars in young stellar clusters show an apparent ``luminosity problem", where their luminosity is observed to be about an order of magnitude lower as compared to that predicted by theoretical models (Kenyon et al. 1990).
Variable young stars called FUors (prototype- FU Orionis) indeed show brightening by a factor of $\sim100$, which can last for several decades (Hartmann \& Kenyon 1996).  
Magneto-rotational instability (MRI) and  gravitational instability (GI) are two of the prominent mechanisms for angular momentum transport in a  protoplanetary disk.
Gammie (1996) suggested that the midplane temperature of a protoplanetary disk at $\sim1$ au is low enough to suppress the MRI, and the accretion occurs only through a thin surface layer due to cosmic ray ionization.
Armitage et al. (2001) demonstrated that the transport of gas due to GI in the outer regions may lead to pile up of material at the magnetically inactive dead zone. 
The increased viscous heating in the this zone can eventually trigger MRI if the midplane temperature reaches a critical value ($T_{\rm crit}$) when the alkali metals are ionized, giving rise to the gravo-magneto (GM) instability in the disk.


\section{Results}

We performed global hydrodynamic simulations of protoplanetary disks in the thin disk limit, including the gravitational collapse phase of the cloud core (for details about the hydrodynamic code, see Vorobyov \& Basu 2015). 
The kinematic viscosity was parametrized using a variable {Shakura \& Sunyaev (1973)} $\alpha$ parameter-   
$\alpha_{\rm eff}=  (\Sigma_a \alpha_a + \Sigma_d \alpha_d)/ (\Sigma_a + \Sigma_d),$
where the subscript `a' denotes the gas surface density of the MRI active layer, while `d' denotes that of the dead zone.
This prescription of the angular momentum transport is consistent with the expectations from a layered disk model.

\begin{figure}
\begin{center}
\includegraphics[width=0.995\textwidth]{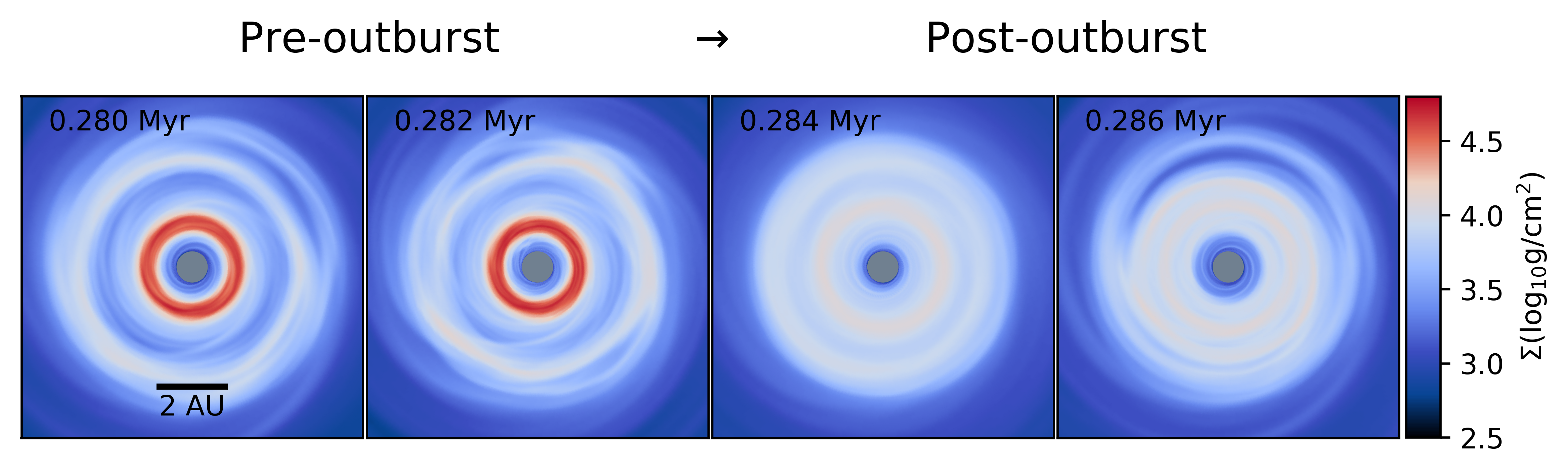}
\includegraphics[width=\textwidth]{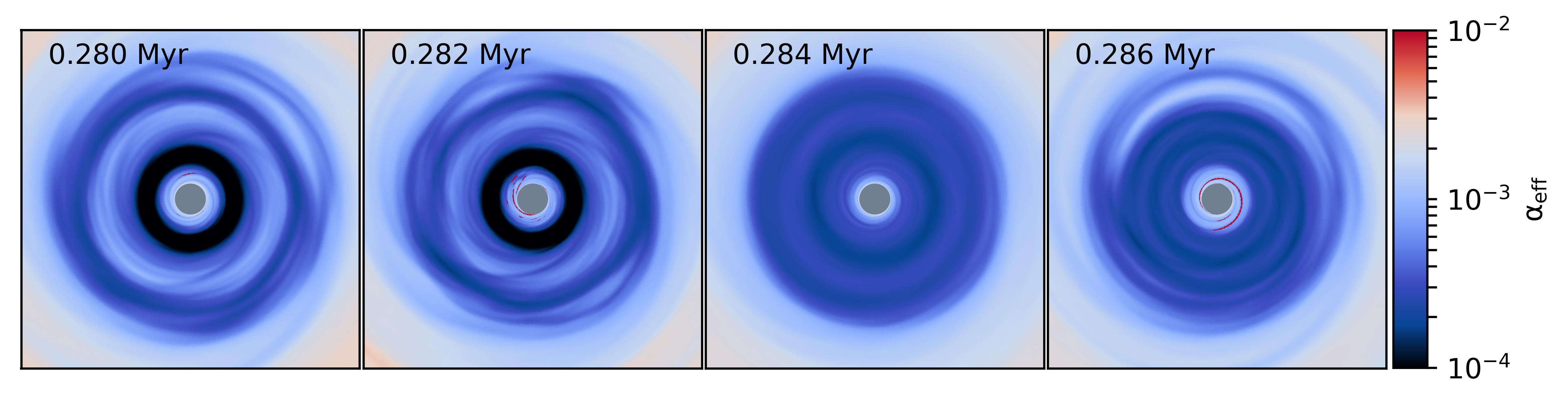}
\end{center}
\vspace{-0.1cm}
\caption{Evolution of the surface density and effective $\alpha$-parameter in the inner 10 au box, showing the progression of the dead zone across an outburst.}
\label{fig1}
\end{figure}

With the canonical values of layered disk parameters ($T_{\rm crit}=1300 {\rm K}, \Sigma_{\rm a}=100$ $ {\rm g/cm^2}$), self-consistent dead zone formed with the variable $\alpha$ prescription; while such zone did not form in (an otherwise identical) fully MRI active disk. 
{The dead zone showed formation of high surface density and low effective viscosity rings between $1-5$ au} (Figure \ref{fig1}, first two columns).

\begin{figure}
\begin{center}
\includegraphics[width=0.8\textwidth]{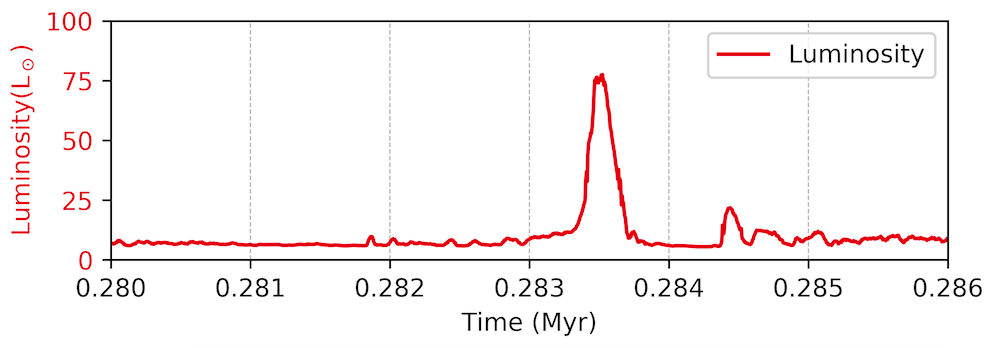}
\end{center}
\vspace{-0.1cm}
\caption {Total luminosity of the outbursting system, depicting an eruption similar to FUors.}
\label{fig2}
\end{figure}

Since the dead zone can not effectively transport angular momentum, the gas continued to accumulate in its vicinity from the outer regions. The central disk temperature eventually reached $T_{\rm crit}$ through viscous heating triggering MRI, and thus the GM instability. 
The dead zone disappeared (see Figure 3, last two columns) and the accumulated material was accerted onto the star.
An associated eruption was observed, which can be considered as FUor-like outburst. 
The eruption typically lasted few hundred years, while the total luminosity of the system increased to $\sim100 L_{\odot}$ (Figure \ref{fig2}).\\

\end{document}